\documentstyle[epsfig, twocolumn]{iso98}

\newcommand{\thepapername}{}

\newcommand{\Frad}{F_{\rm rad}}
\newcommand{\Fpe}{F_{\rm pe}}
\newcommand{\Fpd}{F_{\rm pd}}
\newcommand{\Fdrag}{F_{\rm drag}}
\newcommand{\Ftot}{F_{\rm tot}}
\newcommand{\vd}{v_{\rm drift}}
\newcommand{\vdi}{v_{\rm drift,i}}
\newcommand{\Fstop}{F_{\rm stop}}
\newcommand{\simgt}{\lower.5ex\hbox{$\; \buildrel > \over \sim \;$}}

\begin{document}

\setlength{\parindent}{0pt}
\setlength{\parskip}{ 10pt plus 1pt minus 1pt}
\setlength{\hoffset}{-1.5truecm}
\setlength{\textwidth}{ 17.1truecm }
\setlength{\columnsep}{1truecm }
\setlength{\columnseprule}{0pt}
\setlength{\headheight}{12pt}
\setlength{\headsep}{20pt}
\pagestyle{veniceheadings}

\title{\bf GRAIN DYNAMICS IN PHOTODISSOCIATION REGIONS}

\author{{\bf J.C.~Weingartner$^1$, B.T.~Draine$^2$} \vspace{2mm} \\
$^1$Physics Department, Princeton University, Princeton, NJ, USA \\
$^2$Princeton University Observatory, Princeton, NJ, USA}

\maketitle

\begin{abstract}

We discuss the forces on grains exposed to anisotropic radiation
  fields, including the usual ``radiation pressure'' force and also
  recoil forces due to photoemitted electrons, photodesorbed hydrogen
  atoms, and hydrogen molecules which form on the grain surface.  We
  show that these forces can lead to grain dynamics in
  photodissociation regions which result in enhanced dust-to-gas
  ratios.  Since the gas heating is probably dominated by
  photoelectric emission from dust, this might explain the unusually
  high gas temperatures inferred from ISO observations of molecular
  hydrogen in photodissociation regions.  
  \vspace {5pt} \\


  Key~words: dust; photodissociation regions.

\end{abstract}

\section{INTRODUCTION}

Far ultraviolet radiation incident on a cloud of molecular material
dissociates molecules, giving rise to a surface layer of largely
atomic material, called a photodissociation region (PDR).  When a hot
star is located near a molecular cloud, much of the star's radiant
output is absorbed and reprocessed in the resulting PDR.  Dust in the PDR
efficiently absorbs UV photons, which results in thermal IR emission.
The warm gas cools via forbidden transitions in metal atoms and ions
and via rovibrational transitions in H$_2$.
Thus, the conditions in PDRs are probed
observationally by IR studies.  Many theoretical PDR models have been
developed, and include quantitative determinations of IR diagnostics
(e.g. Tielens \& Hollenbach 1985; Draine \& Bertoldi 1996; see also
the review by Hollenbach \& Tielens 1997 and references therein).  The
detailed understanding of PDRs will yield information on the structure
of molecular clouds and their interaction with young stars, and thus
ultimately on the process of star formation.

As discussed by Draine \& Bertoldi
in this volume, observations with the ISO short 
wavelength spectrometer have yielded the H$_2$ rotational distribution
functions in PDRs, from which surprisingly high values of gas
temperature (500 -- 1000~K) have been inferred.  It is not yet clear
what heating processes are able to maintain the gas at such high
temperatures.  Photoelectric emission from dust grains is expected to
be a major mechanism, but calculated heating rates are inadequate.
Here we consider the possibility of enhanced dust-to-gas ratios in
PDRs, which would imply increased photoelectric heating rates.

\section{GRAIN DYNAMICS}

Enhanced dust-to-gas ratios can result when grains drift with respect
to the gas.  We will begin by describing the motion of the gas (see
Figure 1).  Ionizing radiation from the hot star located near the
molecular cloud (MC) photoevaporates material at the cloud surface,
which then expands into the intercloud medium in a photoevaporative
flow (PF).  Thus, the ionization front propagates into the cloud.  In
our analysis, we adopt the frame of reference in which the ionization
front is stationary; in this frame material flows through the PDR.

\begin{figure}[!h]
  \begin{center}
  \leavevmode
  \centerline{\epsfig{file=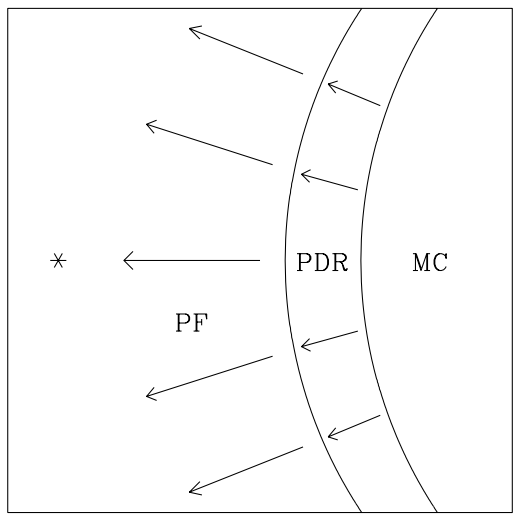,width=8.0cm}}
  \end{center}
  \caption{\em The physical picture.}
\end{figure}

If no other forces acted on the grains, they would simply be dragged along
by the gas.  However, there are forces associated with the anisotropic
stellar UV radiation field.  In addition to the usual ``radiation
pressure'' force due to absorption and scattering ($\Frad$), there are
also recoil forces, due to photoemitted electrons ($\Fpe$) and due to
photodesorbed adatoms and H$_2$ molecules which form on the grain
surface ($\Fpd$).  As a consequence of these ``radiative'' forces, the
grains drift with respect to the gas, at speed $\vd$.  If
the gas moves at speed $v_0$ with respect to the ionization front then
the grains move at speed $v_0 - \vd$.  Deep within the PDR, where the
stellar UV has been attenuated, $\vd = 0$.  The grain overdensity near
the ionization front is obtained using the continuity equation:
\begin{equation}
\frac{n_{\rm i}}{n_0} = \left(1-\frac{\vdi}{v_0}\right)^{-1},
\end{equation}
where $n_0$ ($n_{\rm i}$) is the grain number density deep within the
PDR (near the ionization front) and $\vdi$ is the drift speed near the
ionization front; we assume one-dimensional flow and ignore magnetic fields.
Since the drag force is roughly proportional to the
drift speed, the overdensity is also given by
\begin{equation}
\frac{n_{\rm i}}{n_0} = \left(1-\frac{\Ftot}{\Fstop}\right)^{-1},
\end{equation}
where $\Ftot$ is the sum of the radiative forces near the ionization
front and $\Fstop$ is the magnitude of the radiative force which would
result in $\vd = v_0$ (in which case the grains would be stopped near
the ionization front).  

\section{EVALUATION OF RADIATIVE FORCES}

Throughout, we assume spherical grains, so that Mie theory can be
applied for the grain optics.  

\subsection{Radiation Pressure Force}

The radiation pressure force is given by 
\begin{equation}
\Frad = \pi a^2 \int_0^{\nu_{\rm H}} 
d\nu \, u_{\nu} \left[Q_{\rm abs} + Q_{\rm sca} (1-
\langle\cos\theta\rangle)\right],
\end{equation}
where $a$ is the grain radius, $\nu_{\rm H} = 13.6 \, {\rm eV} / h$, 
$u_{\nu}$ is the radiation energy
density per frequency interval, and $\langle\cos\theta\rangle$ is the
usual scattering asymmetry factor.  We compute the absorption and
scattering efficiency factors, $Q_{\rm abs}$ and $Q_{\rm sca}$
respectively, using a Mie theory code derived from BHMIE (Bohren \&
Huffman 1983) with dielectric functions as described by Draine \& Lee
(1984) and Laor \& Draine (1993).

\subsection{Photoelectric Force}

We adopt a physical model for the photoelectric emission
process (see Bakes \& Tielens 1994, Weingartner \& Draine 1999a).  
Of course, the
photoemission rate and the total energy of the emitted electrons
depend on the grain's ionization potential, $IP$; for simplicity we
take the electron energy $E_{\rm el} = (h\nu - IP)/2$, where $h\nu$ is
the energy of the absorbed photon.  The ionization potential depends
on the grain charge; the charge distribution is determined by the
rates of electron attachment by accretion from the gas and removal by
photoemission.  Thus, the average photoelectric force depends on the
ambient conditions; including the spectrum and energy density of the
radiation and the gas density, ionization, and temperature; through
their effects on the charge distribution.  For a fixed charge state, 
the force is given by
\begin{equation}
\Fpe = \pi a^2 \int_{\nu_{\rm Z}}^{\nu_{\rm H}} d\nu \, \frac{c u_{\nu}}{h\nu} 
A Y Q_{\rm abs} S \sqrt{2 m_e E_{\rm el}},
\end{equation}
where $\nu_{\rm Z} = IP / h$, $Y$ is the photoelectric yield, the 
emission asymmetry factor $A(h\nu,a)$ measures the asymmetry
in the emission of photoelectrons over the grain surface, the recoil
suppression factor $S$ accounts for electron emission in directions
other than the surface normal, and $m_e$ is the electron mass.

We follow the simple prescription of Kerker \& Wang (1982) for determining
the asymmetry factor $A$.  
The probability of photoemission from any site on the surface is taken
to be proportional to the electric intensity ${\left| \bf{E} \right|}^2$ 
just below the surface at that point.  Thus, 
\begin{equation}
A(h\nu, a) = \frac{-\int_0^{\pi} \sin\theta \cos\theta{\left| \bf{E}(\theta)
 \right|}^2 \, d\theta}{\int_0^{\pi} \sin\theta {\left| \bf{E}(\theta)
\right|}^2 \, d\theta},
\end{equation}
where $\theta$ is the polar angle with respect to the direction of the 
incident radiation.  In Figure 2 we display $A(h\nu, a)$
as a function of incident photon energy for graphite grains of various 
sizes.  

\begin{figure}[!h]
  \begin{center}
  \leavevmode
  \centerline{\epsfig{file=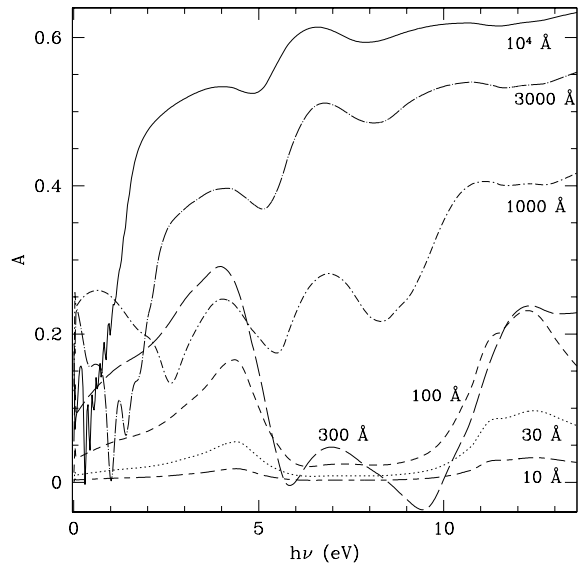,width=8.0cm}}
  \end{center}
  \caption{\em The emission asymmetry factor $A$ as a function of the
           incident photon energy $h \nu$, for graphite grains.  The grain
	   radius is indicated for each curve.}
\end{figure}

In determining the recoil suppression factor $S$ we assume that the 
electrons emerge symmetrically with respect to the local surface normal, 
with a ``cosine-law'' angular distribution (i.e.\ the emission rate at
angle $\psi$ with respect to the surface normal $\propto \sin\psi \cos\psi$).
For an uncharged grain, this 
would imply $S=2/3$.  When the grain is charged, the electron escapes
on a hyperbolic trajectory, so that when it is at infinity
its velocity vector makes an angle $\psi_{\infty}$ with respect to
the surface normal which differs from the corresponding angle at the 
surface, $\psi$.  See Weingartner \& Draine (1999b) for the detailed
calculation.

\subsection{Photodesorption Force}

In our simple model, we consider the exchange of hydrogen between the
gas phase and the grain surface.
We assume that the surface is entirely covered with chemisorption sites,
as would be the case for bare graphite or silicate grains without ice
mantles.  When a gas phase H atom strikes the grain surface, it
arrives either at an empty site or at a site which is already occupied by an
adsorbed H atom.  In the former case, it occupies the site with some
sticking probability; otherwise it reflects and remains in the gas phase.
In the latter case, it recombines with the resident H with some
recombination probability; otherwise it reflects and remains in the gas
phase.  We assume that H$_2$ formed on the grain surface is
immediately ejected from the grain.  Adatoms are removed from the
grain surface via recombination and photodesorption.  

Suppose that the grain's spin axis is aligned with the direction of
the radiation anisotropy.  Since the photodesorption rate is greater in the
illuminated hemisphere than in the non-illuminated hemisphere,
relatively more H leaves the surface as atoms in the former and as
molecules in the latter.  A net recoil force is expected, for two
reasons.  First, the total number of particles leaving the illuminated
hemisphere exceeds that for the non-illuminated hemisphere.  Second,
the outgoing atoms likely carry a different momentum than the outgoing
molecules.  A net recoil force of zero is possible, but only if the
momenta and sticking and recombination probabilities conspire to have
the right values.  The photodesorption force is greater when the spin
axis makes a non-zero angle with respect to the radiation direction,
since H which arrives at non-illuminated sites can be transported to
illumination and photodesorbed.  For a fuller discussion of the
adopted values for relevant parameters and the details of the
calculation, see Weingartner \& Draine (1999b).

\section{THE S140 PDR}

Timmermann et~al. (1996) studied the PDR associated with the S140 H$\,$II
region using ISO and concluded that the gas temperature is $\simgt
500$~K.  We have evaluated the radiative forces for graphite 
and silicate grains
and the conditions in this PDR.  To determine $\Fstop$, we employed
the expression for $\Fdrag$ in Draine \& Salpeter (1979) and evaluated
$v_0$ using a highly simplified model for the flow in the PDR; see
Weingartner \& Draine (1999b) for details.  We display the ratios of
the radiative forces to $\Fstop$, as a function of grain size, in Figures
3 and 4.  

\begin{figure}[!h]
  \begin{center}
  \leavevmode
  \centerline{\epsfig{file=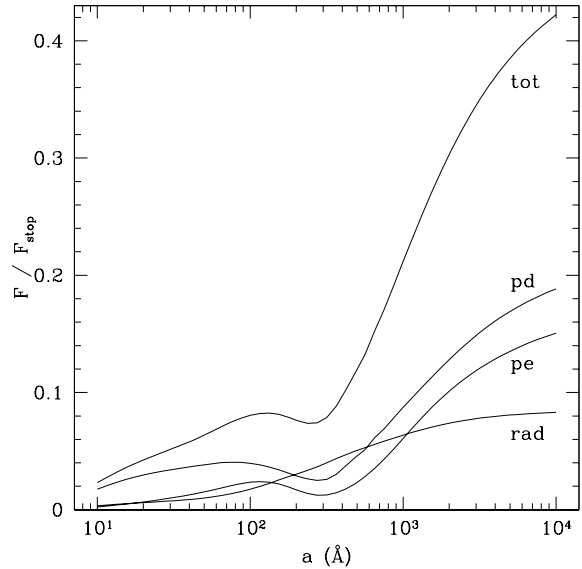,width=8.0cm}}
  \end{center}
  \caption{\em Ratios of the radiative forces to $\Fstop$, the force
  needed to stop graphite grains relative to the ionization front,
  near the ionization front.}
\end{figure}

\begin{figure}[!h]
  \begin{center}
  \leavevmode
  \centerline{\epsfig{file=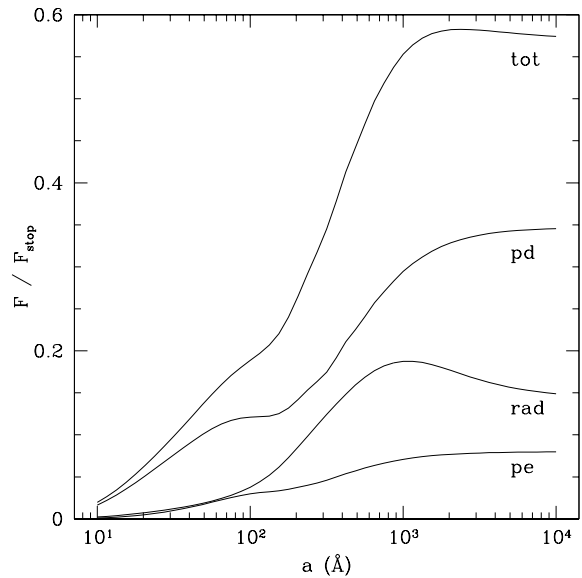,width=8.0cm}}
  \end{center}
  \caption{\em Same as Figure 3, but for silicate grains.}
\end{figure}

\section{CONCLUSIONS AND FUTURE WORK}

For the S140 PDR, the forces associated with the anisotropic radiation are 
apparently large enough to result in significant grain overdensities.  
It is important to note, however, that the total radiative force is highly
uncertain, due to uncertainties in the microphysics of the
photoemission and photodesorption processes.  

Our program of study will include an examination of more PDRs and also
simulations of the dynamics which result when $\Ftot > \Fstop$; i.e.\
when the grains near the ionization front are actually pushed deeper
into the PDR.  In this case, the further upstream a grain is pushed,
the less effective the push, because the radiation is attenuated by
other grains downstream.  Also, we will add magnetic fields to the analysis.
Finally, of course, we will estimate the implications for gas heating.

\section*{ACKNOWLEDGMENTS}

This research was supported in part by NSF grants AST-9219283 and 
AST-9619429 and by an NSF Graduate Fellowship to JCW.  We are grateful to
R.H. Lupton for the availability of the SM plotting package.

\end{document}